\RequirePackage{etoolbox}
\newbool{ipcosubm}
\setbool{ipcosubm}{false} 

\ifbool{ipcosubm}{%
    \documentclass[runningheads,envcountsame,orivec]{llncs}
}{%
    \documentclass{article}
}
\usepackage{graphicx} 

\ifbool{ipcosubm}{%
    \usepackage[T1]{fontenc}
    \usepackage{lmodern}
    \PassOptionsToPackage{style=ext-numeric-comp,sorting=nyt,articlein=false}{biblatex}
}{%
    \PassOptionsToPackage{style=alphabetic,citestyle=alphabetic,maxalphanames=6}{biblatex}
}
\usepackage[
backend=biber,
maxcitenames=999,
mincitenames=999,
maxbibnames=99,
giveninits=true,
url=false,
doi=true,
isbn=true,
backref=true
]{biblatex}

\ifbool{ipcosubm}{%
    \DeclareFieldFormat{title}{#1}
    \DeclareFieldFormat[article,inproceedings,incollection,inbook]{title}{#1}
    \DeclareFieldFormat{journaltitle}{#1}
    \DeclareFieldFormat{booktitle}{#1}
    \DeclareFieldFormat{pages}{#1}
    \DeclareFieldFormat[article]{volume}{\textbf{#1}}
    \DeclareFieldFormat[article]{number}{(#1)}
    
}{%
    \usepackage{amsthm}
}
\usepackage{amsfonts,amsmath,amssymb}
\usepackage{mathtools}
\usepackage{enumerate,graphicx}
\usepackage{xspace}
\usepackage{boxedminipage}
\usepackage{url}
\usepackage{hyperref}
\hypersetup{
    colorlinks=true,
    linkcolor=violet,
    filecolor=magenta,      
    urlcolor=cyan,
    citecolor=blue,
    pdffitwindow=true,
}
\usepackage{soul}
\usepackage{accents}
\usepackage{multirow}
\usepackage[ruled,vlined]{algorithm2e}
\usepackage{booktabs}
\usepackage{enumitem}
\ifbool{ipcosubm}{}{\usepackage{fullpage}}
\usepackage{tikz}
\usepackage{tikz-cd}
\ifbool{ipcosubm}{%

    \spnewtheorem{claim}[theorem]{Claim}{\bfseries}{\itshape}
    \spnewtheorem*{theorem*}{Theorem}{\bfseries}{\itshape}
    \spnewtheorem{fact}{Fact}{\bfseries}{\itshape}
    \spnewtheorem{assumption}[theorem]{Assumption}{\bfseries}{\itshape}
    \spnewtheorem{prop}[theorem]{Proposition}{\bfseries}{\itshape}
    \spnewtheorem{observation}[theorem]{Observation}{\bfseries}{\upshape}
}{}
\usepackage{thmtools}
\usepackage{thm-restate}
\ifbool{ipcosubm}{%
    \let\lncsproof\proof
    \let\endlncsproof\endproof
    \renewenvironment{proof}[1][Proof]
        {\def\proofname{#1}\lncsproof}{\qed\endlncsproof}
}{}
\usepackage{needspace}
\usepackage{array}
\usepackage{booktabs}
\usepackage{threeparttable}
\usepackage{tcolorbox}
\usepackage{ulem}
\usepackage{xfrac}
\ifbool{ipcosubm}{}{\usepackage{subcaption}}
\usepackage[capitalize, nameinlink]{cleveref}
\AddToHook{cmd/appendix/before}{%
    \crefalias{section}{appendix}%
}
\crefname{claim}{Claim}{Claims}
\Crefname{claim}{Claim}{Claims}
\crefname{observation}{Observation}{Observations}
\Crefname{observation}{Observation}{Observations}
\DeclareMathOperator*{\argmax}{arg\,max}

\DeclareMathOperator{\head}{head}
\DeclareMathOperator*{\surplus}{surplus}

\usepackage{color-edits}[showdeletions]

\newcommand{\OPT}{\mathrm{OPT}}

\ifbool{ipcosubm}{}{%
\newtheorem*{theorem*}{Theorem}

\newtheorem{lemma}{Lemma}
\newtheorem{theorem}[lemma]{Theorem}

\newtheorem{claim}[lemma]{Claim}

\theoremstyle{definition}

\newtheorem{definition}[lemma]{Definition}
}

 \usepackage{booktabs}
 \usepackage{tabularx}
 \usepackage{array}

\newcolumntype{Y}{>{\raggedright\arraybackslash}X}

\title{Improved Approximation for Weighted \texorpdfstring{$k$}{k}-Set Packing via a Simple Two-Phase Approach}
\ifbool{ipcosubm}{%
\titlerunning{Improved Approximation for Weighted \texorpdfstring{$k$}{k}-Set Packing}
\author{}
\authorrunning{}
\institute{}
\hypersetup{pdfauthor={}}
}{%
\author{%
Mehrshad Taziki\thanks{
ETH Zurich, Zurich, Switzerland.
Email: \href{mailto:mtaziki@ethz.ch}%
{mtaziki@ethz.ch}.}
\and
Rico Zenklusen\thanks{
ETH Zurich, Zurich, Switzerland.
Email: \href{mailto:ricoz@ethz.ch}%
{ricoz@ethz.ch}.}
}
}
\date{}

\begin{document}

\maketitle

\begin{abstract}
Weighted $k$-set packing asks for a maximum-weight collection of pairwise disjoint sets, each of cardinality at most $k$.
For every fixed $k\in\mathbb Z_{\geq1}$, we give a polynomial-time algorithm with approximation factor $\bigl(1-\sfrac1{\sqrt3}\bigr)k+O(\log k) \approx 0.4226k + O(\log k)$, improving the previous best factor of $0.4621k + O(\sqrt{k})$ through a simple two-phase approach.
In a first phase, our algorithm constructs a packing by processing random weight buckets of varying widths with an unweighted packing algorithm.
We show that a small expected weight of this packing relative to the optimum forces a large expected total \emph{surplus} in the optimum.
Surplus quantifies the potential for improving a packing through \emph{star swaps}.
In the second phase, we solve and round a configuration LP to turn sufficiently large surplus into an increase in expected weight.
Our procedure can easily be derandomized.
\ifbool{ipcosubm}{%
}{}
\end{abstract}

\section{Introduction}
\label{sec:introduction}

Weighted $k$-set packing is a central packing problem in combinatorial optimization. Given a family of sets of cardinality at most $k$, the goal is to select pairwise disjoint sets of maximum total weight. The problem generalizes maximum weight matching in graphs to hypergraphs and provides a basic model for optimization under disjointness constraints. Its approximability has been studied extensively for decades, with local search and linear programming playing prominent roles~\cite{hurkens_schrijver1989,berman2000,chan_lau2012,cygan_unweighted}. Determining the best approximation factor achievable as a function of $k$ remains a fundamental open question.

Formally, an instance of weighted $k$-set packing, for some $k\in \mathbb{Z}_{\geq 1}$, consists of a finite ground set $U$, a family $E\subseteq 2^U$ of sets, and positive rational weights $w\colon E\to\mathbb Q_{>0}$. Each $e\in E$ has an element set $v(e)\subseteq U$ of cardinality between one and $k$. A \emph{packing} is a subfamily $F\subseteq E$ whose element sets are pairwise disjoint, and its weight is $w(F)\coloneqq\sum_{e\in F}w(e)$. We seek a packing of maximum weight and write $\OPT\subseteq E$ for an optimal packing. Smaller sets can be padded with private dummy elements, so allowing cardinality at most $k$ is equivalent to requiring cardinality exactly $k$.
The problem includes $k$-dimensional matching, where $U$ is partitioned into $k$ parts and every input set contains exactly one element from each part. All approximation guarantees for $k$-set packing therefore also apply to this special case.

The natural greedy algorithm repeatedly selects a heaviest remaining set and discards every set that intersects it.
This gives a $k$-approximation: each selected set can exclude at most $k$ disjoint optimum sets, each of no greater weight.
This guarantee is tight, since greedy may select a single set instead of $k$ mutually disjoint sets of almost the same weight.
Going beyond this basic factor spawned a long line of research on local search algorithms, which iteratively improve a packing by replacing some of its sets with disjoint sets of greater total weight.

For the unweighted case, where each set has weight one, \textcite{hurkens_schrijver1989} obtained a $(\frac{k}{2}+\varepsilon)$-approximation through exchanges of constant size. Their argument also yields this factor for maximum independent set in $(k+1)$-claw-free graphs, i.e., graphs in which no vertex has $k+1$ pairwise nonadjacent neighbors. Set packing reduces to this problem via its conflict graph, whose vertices are the input sets and whose edges join distinct intersecting sets.
\textcite{cygan2013sell} achieved a $(\frac{k+1}{3}+\varepsilon)$-approximation in quasi-polynomial time through swaps of size $O(\log |E|)$.
\textcite{cygan_unweighted} later got the same guarantee in polynomial time by using swaps of bounded pathwidth.
\textcite{furer_unweighted} subsequently obtained the same factor with improved running time. These guarantees hold for every fixed $k \geq 2$ and $\varepsilon>0$. The unweighted algorithms exploit the number and incidence structure of conflicting sets, information that is less directly useful when their weights differ substantially.

For general weights, \textcite{chandra_halldorsson2001} obtained the first asymptotic improvement over the greedy factor $k$, achieving a $(\frac{2(k+1)}{3}+\varepsilon)$-\hspace{0pt}approximation for every fixed $k$ and $\varepsilon>0$. Their algorithm starts from a greedy packing and repeatedly chooses a local exchange maximizing the ratio of inserted to deleted weight.
These exchanges are similar to what we call star swaps in \cref{sec:overview}.
Subsequently, \textcite{berman2000} introduced non-oblivious local search that optimizes the sum of squared weights and achieves a $(\frac{k+1}{2}+\varepsilon)$-approximation.
\textcite{neuwohner2021} gave the first improvement over this guarantee, obtaining a $(\frac{k+1}{2}-\frac{1}{63700992}+\varepsilon)$-approximation for every fixed $k\in\mathbb Z_{\geq2}$ and $\varepsilon>0$. By analyzing larger squared-weight exchanges, \textcite{thiery_non_oblivious} obtained a $(\frac{k+1}{2}-0.214)$-approximation for $k\geq3$, giving a factor of $1.786$ for $k=3$, and a sharper asymptotic guarantee of $\frac{k}{2}+\frac{1}{6}+o(1)$ as $k\to\infty$. The guarantees of all four algorithms hold more generally for maximum weight independent set in $(k+1)$-claw-free graphs.

A limitation of this approach is that, for every fixed power of the weight function, exchanges of logarithmically bounded size can have a locality gap approaching $\frac{k}{2}$~\cite{meike_non_oblivious}. This obstruction already occurs in set-packing instances. \textcite{neuwohner2023} surpassed this threshold by combining non-oblivious local search with unweighted packing algorithms, obtaining a $0.4986k+O(1)$ approximation.

The best prior guarantee for large $k$ is due to \textcite{singer2026_submodular_maximization}, who obtained an approximation factor of $\frac{\ln4}{3}k+O(\sqrt{k}) \approx 0.4621 k + O(\sqrt{k})$ in expectation.
Their algorithm processes weight buckets with a random logarithmic shift and fixed multiplicative width two.
They alternate between selecting sets within a bucket using an unweighted packing algorithm and applying ordered local search to all sets processed so far. The local search revisits earlier choices to repair the losses caused by greedy processing of the buckets.

\paragraph{Our result.}
We obtain the following stronger asymptotic guarantee for weighted $k$-set packing with a simple two-phase approach.
For simplicity of presentation, and to better highlight our main algorithmic ideas, we first describe a randomized algorithm obtaining the desired guarantees in expectation.
\cref{sec:derandomization} then shows how it can easily be derandomized.

\begin{restatable}[Main result]{theorem}{mainapproximation}
\label{thm:main}
For every fixed $k\in \mathbb{Z}_{\geq 1}$, there is a randomized polynomial-time algorithm for weighted $k$-set packing that returns a packing $C$ satisfying
\[
    w(\OPT)\leq\left(\left(1-\frac1{\sqrt3}\right)k+O(\log k)\right)
        \mathbb E[w(C)].
\]
The expectation is over the internal randomness of the algorithm.
\end{restatable}

Compared to the previous best guarantee of \textcite{singer2026_submodular_maximization}, our leading constant is $1-\frac{1}{\sqrt3}\approx 0.4226$ instead of $\frac{\ln 4}{3}\approx 0.4621$.
Moreover, the lower-order term is also improved from $O(\sqrt{k})$ to $O(\log k)$.

\paragraph{Techniques behind the improvement and organization of the paper.}
As in \textcite{singer2026_submodular_maximization}, we use an unweighted packing algorithm within random weight buckets. Their procedure interleaves bucket processing with repeated weighted local search over all sets processed so far. We first complete the greedy construction using new distributions with varying bucket widths, then solve and round a single configuration LP over what we call star swaps around the resulting packing.
The analysis of \textcite{chandra_halldorsson2001} also groups optimum sets by their heaviest neighbors to share deletion costs. Their exchanges allow the inserted sets to have any common neighbor in the current packing; our LP restricts them to share a heaviest neighbor in the initial packing. Moreover, we recover improvements by rounding this LP relative to a fixed packing, whereas their algorithm repeatedly updates the packing through greedy local exchanges.
A key new ingredient that makes our approach work is a surplus guarantee that links the phases.
Surplus measures the weight available to pay for shared deletion costs in star swaps around a given packing.
We show that if the greedy construction returns a packing of small expected weight, then the optimum must have a large expected total surplus.

To make this connection precise, \cref{sec:overview} introduces surplus and star swaps and states the two phase guarantees. \Cref{sec:greedy-packing} develops the random bucketing and greedy construction, while \cref{sec:star-lp} presents the configuration LP and its rounding. The respective phase guarantees are proved in \cref{sec:first-phase-analysis} and \ifbool{ipcosubm}{\cref{app:surplus-recovery}}{\cref{sec:star-lp}}; \cref{sec:combining-phases} combines them to establish \cref{thm:main}. Proofs of the bucket distribution properties and the required unweighted packing guarantee appear in \cref{app:bucket-properties,app:cygan-guarantee}, respectively.
\Cref{sec:derandomization} shows how to derandomize our algorithm.

\paragraph{Further context.}
LP relaxations have also been studied for hypergraph matching. For $k\in \mathbb{Z}_{\geq2}$, \textcite{chan_lau2012} gave an algorithmic analysis of the standard relaxation with factor $k-1+\frac{1}{k}$, improving to $k-1$ for weighted $k$-dimensional matching, and studied stronger relaxations and their connections to local search. Our LP instead chooses improvements around an already constructed packing. For small $k$, specialized analyses give stronger guarantees than asymptotic formulas alone suggest; a recent preprint of \textcite{tong_xu2026} gives a $1.6908$-approximation for weighted $3$-set packing, and consequently for weighted $3$-dimensional matching and weighted vertex-disjoint triangle packing.

On the hardness side, \textcite{hazan_safra_schwartz2006} proved an $\Omega(\frac{k}{\log k})$ lower bound even for unit weights, unless $\mathrm P=\mathrm{NP}$. \textcite{lee_svensson_thiery2025} strengthened this to linear hardness, and \textcite{minzer_zheng2026} obtained $\frac{k}{8+\varepsilon}$ hardness for every fixed $\varepsilon>0$ and suitable sufficiently large $k$, under $\mathrm{NP}\not\subseteq\mathrm{BPP}$. These lower bounds already hold for unweighted $k$-dimensional matching, and therefore also hold for $k$-matroid intersection, $k$-matchoids, and matroid $k$-parity. Thus a linear dependence on $k$ is necessary, while the best achievable leading constant remains open.

\section{Overview of the approach}
\label{sec:overview}

For $k=1$, we return an optimal packing of singletons. For $k\geq2$, we use a two-step procedure. The first phase constructs a maximal packing $A$ with a guarantee that also quantifies opportunities for improvement; we do not obtain our final approximation from this phase alone. In the second phase, we exploit these opportunities through \emph{star swaps}. Such a swap removes a set $h\in A$, called its \emph{head}, inserts a nonempty packing $S\subseteq E$ whose members all intersect $h$, and removes every other set of $A$ that intersects a member of $S$. Necessarily, $|S|\leq k$. The main idea is to pay for removing $h$ only once, allowing several insertions to share this cost.

\paragraph{Surplus measures what can pay for the head.}
For $e\in E$ and $F\subseteq E$, let $\mathcal N(e,F)\coloneqq\{f\in F\colon v(f)\cap v(e)\neq\emptyset\}$, including $e$ itself when $e\in F$. Since $A$ is maximal, $\mathcal N(e,A)$ is nonempty. Let $\head(\cdot\mid A)\colon E\to A$ be the function that maps each set $e$ to a heaviest neighbor of it, with ties broken by a fixed rule. Define $\surplus(\cdot\mid A)\colon E\to\mathbb R_{\geq0}$ by
\[
    \surplus(e\mid A)\coloneqq
    \max\bigl\{w(e)-w\bigl(\mathcal N(e,A)\setminus\{\head(e\mid A)\}\bigr),\,0\bigr\}.
\]
A positive surplus is the weight left from $e$ after paying for all its conflicts except its head. It need not suffice to pay for the head on its own. However, if the members of a star have positive surplus and the same head $h$, the star swap increases weight by at least
\[
    \sum_{s\in S}\surplus(s\mid A)-w(h).
\]
Indeed, after deleting $h$, insert the members one by one. They do not conflict with one another, and each insertion costs at most the weight of its original neighbors other than $h$. Thus each contributes at least its surplus, while $w(h)$ is subtracted only once.

\paragraph{A starting packing with useful surplus.}
The first phase is designed so that a small expected weight of $A$ forces a large expected total surplus in any heavy comparison packing.
Its guarantee is stated below for geometric weights; the reduction from arbitrary positive weights is given in \cref{sec:weight-rounding}.

\Needspace{10\baselineskip}
\begin{restatable}[First-phase guarantee]{theorem}{firstphaseguarantee}
\label{thm:phase1}
For any fixed $k\in \mathbb{Z}_{\geq 1}$, suppose the weights belong to $\{(1+1/k)^j\colon j\in\mathbb Z_{\geq0}\}$, and let $m\coloneqq\left\lceil\log(\sqrt3)/\log(1+1/k)\right\rceil$.
There is a randomized polynomial-time algorithm returning a maximal packing $A$ such that, for every packing $Q\subseteq E$ fixed independently of its randomness, 
\[
    \frac{3\bigl(1+(1+1/k)^{-m}\bigr)}{2+1/k}\,w(Q)
    \leq(k+3)\mathbb E[w(A)]
       +\mathbb E\left[\sum_{e\in Q}\surplus(e\mid A)\right].
\]
The coefficient on the left is $\frac{3+\sqrt3}{2}\bigl(1-O(1/k)\bigr)$. 

\end{restatable}

The approach is based on a new random bucketing of the input sets by weight, followed by creating a packing from the buckets using an unweighted $k$-set-packing algorithm. 
In \cref{sec:greedy-packing}, we first describe the greedy framework, then construct the random buckets with the properties needed for its analysis. The proof of \cref{thm:phase1} is given in \cref{sec:first-phase-analysis}.

For $Q=\OPT$, \cref{thm:phase1} already gives our desired $((1-1/\sqrt3)k+O(\log k))$-approximation when the expected total surplus is $O(\log k)\mathbb E[w(A)]$. It remains to exploit the case in which the surplus term is larger.

\paragraph{Recovering surplus through an LP.}
For $k\in \mathbb{Z}_{\geq2}$ and a maximal packing $A$, the second phase uses a packing linear program with a variable for each star configuration with respect to $A$.
The LP limits the total fractional mass of star members containing any ground-set element to one.
Randomized rounding samples stars in a random order and retains compatible members while controlling the costs of deleted heads. It produces a packing $B$ with the following guarantee.

\begin{restatable}[Surplus recovery]{theorem}{surplusrecovery}
\label{thm:star-lp-rounding}
For any fixed $k\in \mathbb{Z}_{\geq 2}$ and any maximal packing $A$, there is a randomized polynomial-time algorithm returning a packing $B$ such that, for every packing $Q\subseteq E$,
\[
    \mathbb E[w(B)]\geq w(A)
       +\frac1{k+4}\sum_{e\in Q}\surplus(e\mid A)
       -\frac{\ln k}{k+1}\,w(A).
\]
\end{restatable}

Details of the linear program and its rounding are given in \cref{sec:star-lp}. The proof of \cref{thm:star-lp-rounding} appears in \ifbool{ipcosubm}{\cref{app:surplus-recovery}}{\cref{sec:star-lp}}.

\paragraph{Combining the two phases.}

For $k\in \mathbb{Z}_{\geq 2}$, conditioning on the first-phase output $A$ shows that the expected weight increases whenever the expected total surplus exceeds $\frac{k+4}{k+1}\ln k\,\mathbb E[w(A)]$. Thus the surplus that weakens the first guarantee provides the improvement in the second. Returning a heavier packing $C\in\{A,B\}$ gives the claimed $((1-1/\sqrt3)k+O(\log k))$-approximation.
(See \cref{sec:combining-phases} for the details of this argument, which completes the proof of \cref{thm:main}.)

\section{First phase: random bucketing and greedy packing}
\label{sec:greedy-packing}
Before proving \cref{thm:phase1} we describe the greedy procedure for a given partition into weight buckets. This will identify what the bucket boundaries must achieve and motivate their random construction. The empty instance is immediate, so assume $E\neq\emptyset$.

\subsection{The greedy framework}
\label{sec:greedy-framework}
Fix $K\in\mathbb Z_{>0}$ and boundaries $m_0,\ldots,m_K\in\mathbb R_{>0}$ such that $m_0>m_1>\cdots>m_K$, $m_0\geq\max_{e\in E}w(e)$, and $m_K<\min_{e\in E}w(e)$. They partition the sets into buckets
\[
    E_i\coloneqq\{e\in E\colon m_i<w(e)\leq m_{i-1}\},
    \qquad i=1,\ldots,K.
\]
Define $\ell\colon E\to\mathbb R_{>0}$ by $\ell(e)\coloneqq m_i$ for $e\in E_i$, so $\ell(e)$ is the lower boundary of the bucket containing $e$. For now, the boundaries are arbitrary.

Within each bucket, we ignore weights and use an unweighted packing subroutine. We need a maximal packing for which few sets of any comparison packing have only one or two neighbors. The following guarantee follows from the bounded-pathwidth search of \textcite{cygan_unweighted}; for completeness, we give the derivation in \cref{app:cygan-guarantee}.

\begin{theorem}[\cite{cygan_unweighted}]
    \label{thm:cygan-guarantee}
    For any fixed $k\in\mathbb Z_{\geq1}$ and a family $E$ of sets of cardinality at most $k$, there is a deterministic polynomial-time procedure that returns a maximal packing $A\subseteq E$ such that every packing $Q\subseteq E$ satisfies
    \[
        |\{e\in Q\colon |\mathcal N(e,A)|=1\}|\leq|A|
        \qquad\text{and}\qquad
        |\{e\in Q\colon |\mathcal N(e,A)|=2\}|\leq2|A|.
    \]
\end{theorem}

\Cref{alg:greedy-packing} processes the buckets from heaviest to lightest. In bucket $i$, the family $F_i\subseteq E_i$ consists of the \emph{free} sets: those disjoint from all earlier choices. Fix the choices of the unweighted subroutine as a function of its input, so the decision in bucket $i$ depends only on the buckets revealed so far.

\begin{algorithm}[H]
    \caption{Greedy packing by weight buckets}
    \label{alg:greedy-packing}
    \DontPrintSemicolon
    \KwIn{Weight buckets $E_1,\ldots,E_K$, ordered from heaviest to lightest.}
    \KwOut{A maximal packing $A$.}
    $A_{\leq0}\gets\emptyset$\;
    \For{$i=1,\ldots,K$}{
        $F_i\gets\{e\in E_i\colon \mathcal N(e,A_{\leq i-1})=\emptyset\}$\;
        \eIf{$F_i=\emptyset$}{
            $A_i\gets\emptyset$\;
        }{
            Compute a packing $A_i\subseteq F_i$ using \cref{thm:cygan-guarantee}\;
        }
        $A_{\leq i}\gets A_{\leq i-1}\cup A_i$\;
    }
    $A\gets A_{\leq K}$\;
    \KwRet{$A$}\;
\end{algorithm}

The output is a packing, and it is maximal because every set not selected either meets an earlier choice or meets a set selected in its own bucket.

To express the guarantee of this phase, we measure the weight of a comparison set left over after paying for all its neighbors except the heaviest one.
\begin{definition}[Heaviest neighbor]
    \label{def:heaviest-neighbor}
    For a maximal packing $A\subseteq E$, define $\head(\cdot\mid A)\colon E\to A$ by
    \[
        \head(e\mid A)\in\argmax_{f\in\mathcal N(e,A)}w(f),
    \]
    with ties broken by an arbitrary fixed rule.
\end{definition}
\begin{definition}[Surplus]
    \label{def:surplus}
    For a maximal packing $A\subseteq E$, define $\surplus(\cdot\mid A)\colon E\to\mathbb R_{\geq0}$ by
    \[
        \surplus(e\mid A)
        \coloneqq\max\{w(e)+w(\head(e\mid A))-w(\mathcal N(e,A)),\,0\}.
    \]
\end{definition}
Thus, when the surplus is positive, it measures the gain from inserting $e$ after its heaviest neighbor has already been removed. Keeping the cost of the heaviest neighbor separate will allow the second phase to pay for it only once when inserting several disjoint sets with the same heaviest neighbor.

\paragraph{What should the bucket boundaries achieve?}
Fix a comparison packing $Q\subseteq E$. (We will later set $Q$ to be an optimal packing.) For each $e\in Q$, we compare $3\ell(e)$ with
\[
    w(\mathcal N(e,A))+\surplus(e\mid A).
\]
Our goal is to bound $3\sum_{e\in Q}\mathbb E[\ell(e)]$ from above by the expected sum of these quantities, plus $3\mathbb E[w(A)]$.

In each bucket, \cref{thm:cygan-guarantee} ensures that all but at most $3|A_i|$ sets of $Q\cap F_i$ have at least three neighbors in $A_i$. For each such set $e$, these neighbors have total weight at least $3\ell(e)$, so already $w(\mathcal N(e,A))\geq3\ell(e)$. This explains the factor $3$. The remaining free sets may fall short, but their total shortfall in neighborhood weight plus surplus will be bounded above by $3w(A)$ in the analysis.

The boundaries must serve two purposes. First, $\mathbb E[\ell(e)]$ should be bounded below by a large, fixed fraction of $w(e)$ for every fixed set $e\in E$. This lets us turn an upper bound on $\sum_{e\in Q}\mathbb E[\ell(e)]$ into an upper bound on $w(Q)$. Second, we must handle sets blocked by an earlier choice, which never enter the unweighted subroutine. Suppose a set $e\in Q$ is blocked by a set selected before its own bucket is processed. Let $M\in\mathbb R_{>0}$ be the lower boundary of the first bucket in which such a blocking set is selected. The selected set that blocks $e$ has weight at least $M$, and the definition of surplus gives
\[
    w(\mathcal N(e,A))+\surplus(e\mid A)
    \geq w(e)+w(\head(e\mid A))
    \geq w(e)+M.
\]
Thus, it suffices to show that the conditional expectation of $3\ell(e)$, given the buckets and choices revealed up to this point, is at most $M+w(e)$.
The two goals pull in opposite directions: narrow buckets keep $\mathbb E[\ell(e)]$ close to $w(e)$, whereas the conditional upper bound for a blocked set requires its expected lower boundary to lie well below $w(e)$ when $M$ is close to $w(e)$. We next construct random buckets with both properties.

\subsection{Choosing the random buckets}
\label{sec:random-buckets}
\paragraph{Normalizing and rounding weights}
\label{sec:weight-rounding}
Set $r\coloneqq1+1/k\in\mathbb R_{>1}$. Rescale the weights so that their minimum is one, and round each weight down to the largest nonnegative integral power of $r$ that does not exceed it. Each weight decreases by a factor of at most $r$, so an approximation factor for the rounded instance transfers to the original instance with a multiplicative loss of at most $r$. For our $O(k)$ guarantee, this changes the approximation factor by only $O(1)$ additively. From now on, $w$ denotes the rounded weights, and $W\coloneqq\max_{e\in E}w(e)\in\mathbb R_{\geq1}$. Thus every weight lies in $[1,W]$ and on the grid $\{Wr^{-j}\colon j\in\mathbb Z_{\geq0}\}$. 

\paragraph{Constructing the buckets}
Let $m\in\mathbb Z_{>0}$ be the smallest integer such that $r^m\geq\sqrt3$. A bucket of length $i\in\{1,\ldots,m\}$ starting at a boundary $M\in\mathbb R_{>0}$ is the interval $(Mr^{-i},M]$. Its upper and lower boundaries differ by a factor of at most $r^m<r\sqrt3$.

For all buckets after the first, draw the length from a geometric distribution capped at $m$. More precisely, let $G$ be a geometric random variable taking values in $\mathbb Z_{>0}$ with success probability $1/(k+1)$, and let $\nu$ be the law of $\min\{G,m\}$ on $\{1,\ldots,m\}$:
\begin{equation}
    \label{eq:nu-def}
    \nu_i\coloneqq
    \begin{cases}
        (r-1)r^{-i},&1\leq i<m,\\
        r^{-(m-1)},&i=m.
    \end{cases}
\end{equation}
The first bucket uses the law $\pi$ on $\{1,\ldots,m\}$, obtained by conditioning $G$ on $G\leq m$:
\begin{equation}
    \label{eq:pi-def}
    \pi_i\coloneqq\frac{(r-1)r^{-i}}{1-r^{-m}},
    \qquad i=1,\ldots,m.
\end{equation}
The distinction is that $\nu$ puts the entire remaining tail at $m$, whereas $\pi$ conditions on a length of at most $m$. This choice of $\pi$ gives the position of a fixed grid weight relative to the next lower boundary the same distribution at every level of the grid, as stated below.

Set $m_0\coloneqq W$. Draw independent random lengths $X_i\in\{1,\ldots,m\}$ with $X_1\sim\pi$ and $X_i\sim\nu$ for $i\geq2$, and set $m_i\coloneqq m_{i-1}r^{-X_i}$, stopping at the first $K\in\mathbb Z_{>0}$ with $m_K<1$. These boundaries define the buckets $E_i$ and lower boundaries $\ell(e)$ as in \cref{sec:greedy-framework}. Running \cref{alg:greedy-packing} on these buckets completes the first-phase algorithm.

\paragraph{The two boundary properties}
The first property explains why we use $\pi$ for the initial bucket and $\nu$ for all later buckets: Viewing from any fixed grid weight, the number of grid steps to the next lower boundary obeys law $\pi$.
\begin{lemma}[Common boundary distribution]
    \label{lem:idential-dist+expected-ell}
    For every fixed $e\in E$, the ratio $w(e)/\ell(e)$ has distribution $r^X$ with $X\sim\pi$. In particular,
    \begin{equation}
        \label{eq:mean-boundary}
        \mathbb E[\ell(e)]
        =\frac{1+r^{-m}}{r+1}\,w(e)
        =\frac{\sqrt3+1}{2\sqrt3}\left(1-O\left(\frac1k\right)\right)w(e).
    \end{equation}
\end{lemma}
This gives the first property sought in \cref{sec:greedy-framework}: expected lower boundaries are a fixed fraction of the weights. The second property handles blocked sets by controlling the continuation below a revealed boundary, where the remaining lengths are independent draws from $\nu$.

\begin{lemma}[Continuing from a fixed boundary]
    \label{lem:fresh-boundary}
    Starting at a boundary $M\in\mathbb R_{>0}$, generate successive buckets below $M$ using independent lengths distributed according to $\nu$. For $j\in\mathbb Z_{\geq0}$, set $w\coloneqq Mr^{-j}\in\mathbb R_{>0}$ and let $\ell_M(w)\in\mathbb R_{>0}$ be the lower boundary of its bucket. Then
    \[
        3\mathbb E[\ell_M(w)]\leq M+w.
    \]
\end{lemma}

The lemma also applies conditionally on any history that ends at a bucket boundary $M$ and leaves the later lengths unrevealed. In the analysis of the greedy phase, for any set $e$, we condition on the first earlier bucket whose selected packing intersects $e$, and take $M$ to be that bucket's lower boundary. The weight of $e$ together with $M$ then pays for three times the expected lower boundary of $e$. Proofs of both properties are given in \cref{app:bucket-properties}.

\subsection{Analysis of the first phase}
\label{sec:first-phase-analysis}
We can now prove the first-phase guarantee from the overview. 
\firstphaseguarantee*

\begin{proof}
    Fix such a packing $Q$. By \cref{lem:idential-dist+expected-ell}, the left-hand side of the claimed inequality is
    \[
        \frac{3(1+r^{-m})}{r+1}\,w(Q)
        =3\sum_{e\in Q}\mathbb E[\ell(e)].
    \]
    We therefore seek an upper bound on this expected sum in terms of the weight of $A$ and the remaining surpluses. Define $\gamma\colon Q\to\mathbb R_{\geq0}$ by
    \begin{equation}   
        \label{eq:coverage-identity}
        \begin{aligned}
            \gamma(e)&\coloneqq w(\mathcal N(e,A))+\surplus(e\mid A) \\
        &=\max\{w(\mathcal N(e,A)),\,w(e)+w(\head(e\mid A))\}.
        \end{aligned}
    \end{equation}
    Each set of $A$ intersects at most $k$ sets of the packing $Q$, so the neighborhood weights in $\sum_{e\in Q}\gamma(e)$ sum to at most $k\,w(A)$. It therefore suffices to prove
    \[
        3\sum_{e\in Q}\mathbb E[\ell(e)]
        \leq\mathbb E\left[\sum_{e\in Q}\gamma(e)\right]+3\mathbb E[w(A)].
    \]

     We separate the remaining difficulties according to whether $e$ was available when its bucket was processed. Let $F\coloneqq\bigcup_iF_i$, $Q_{\mathrm{free}}\coloneqq Q\cap F$, and $Q_{\mathrm{blocked}}\coloneqq Q\setminus Q_{\mathrm{free}}$. These two subfamilies partition $Q$ for every realization of the buckets. We bound their contributions separately.

    \emph{Free sets.} We show that the cumulative shortfall of $\gamma(e)$ below $3\ell(e)$ is at most $3w(A)$. More precisely, for every realization,
    \begin{equation}
        \label{eq:free-coverage}
        3\sum_{e\in Q_{\mathrm{free}}}\ell(e)
        \leq\sum_{e\in Q_{\mathrm{free}}}\gamma(e)+3w(A).
    \end{equation}
    To prove this, consider a free set $e\in Q\cap F_i$. Maximality of $A_i$ gives a neighbor of $e$ in $A_i$, so $\gamma(e)\geq w(e)+w(\head(e\mid A))\geq2m_i$. If $e$ has at least three neighbors in $A_i$, then $\gamma(e)\geq w(\mathcal N(e,A_i))\geq3m_i$.
    Thus only the sets in
    \[
        S_i\coloneqq\{e\in Q\cap F_i\colon |\mathcal N(e,A_i)|\leq2\}
    \]
    can fall short of $3\ell(e)=3m_i$, and each falls short by at most $m_i$. By \cref{thm:cygan-guarantee}, $|S_i|\leq3|A_i|$. Thus, the total deficit is at most
    \[
        \sum_i m_i|S_i|\leq3\sum_i m_i|A_i|\leq3w(A),
    \]
    where the second inequality follows from $w(e) \geq m_{i}$ for each $e\in A_i$. This establishes \eqref{eq:free-coverage}.

    \emph{Blocked sets.} Here we show that the expected total sum of $\gamma(e)$ suffices without an additive term. Namely,
    \begin{equation}
        \label{eq:blocked-coverage}
        3\mathbb E\left[\sum_{e\in Q_{\mathrm{blocked}}}\ell(e)\right]
        \leq\mathbb E\left[\sum_{e\in Q_{\mathrm{blocked}}}\gamma(e)\right].
    \end{equation}
    Membership in $Q_{\mathrm{blocked}}$ depends on the prior sampled buckets. We therefore fix $e\in Q$ and reveal the buckets from top to bottom; the choice in each bucket only uses the buckets revealed so far. Consider a history $\mathcal H$ ending with a bucket of lower boundary $M$ such that $w(e)\leq M$ and $e$ is blocked for the first time by a set in that bucket. These are precisely the first-blocking histories for $e$. Each such history determines that $e$ is blocked and leaves an independent continuation below $M$ with lengths distributed according to $\nu$.
    Since $w(e)\leq M$ and both lie on the weight grid, $w(e)=Mr^{-j}$ for some $j\in\mathbb Z_{\geq0}$. Thus \cref{lem:fresh-boundary} gives
    \[
        3\mathbb E[\ell(e)\mid\mathcal H]\leq M+w(e).
    \]
    The selected set that blocks $e$ has weight at least $M$, so \eqref{eq:coverage-identity} implies $\gamma(e)\geq M+w(e)$ for every continuation. Averaging over all blocking histories with their original probabilities and summing over $e\in Q$ proves \eqref{eq:blocked-coverage}.

     Since $Q_{\mathrm{free}}$ and $Q_{\mathrm{blocked}}$ partition $Q$, taking expectations in \eqref{eq:free-coverage} and adding \eqref{eq:blocked-coverage} gives the desired bound on $3\sum_{e\in Q}\mathbb E[\ell(e)]$.
\end{proof}

\section{Gaining a fraction of the surplus: star configuration LP}\label{sec:star-lp}
\Cref{thm:phase1} already gives a strong approximation guarantee when the expected total surplus of the comparison packing $Q$ is small relative to $\mathbb E[w(A)]$. For $k\geq2$, we now show that large surpluses provide an opportunity to improve the packing. Starting from any maximal packing $A$, we use a linear program and randomized rounding to construct a packing $B$. The following theorem bounds the expected gain and, in particular, guarantees $\mathbb E[w(B)]>w(A)$ whenever $\sum_{e\in Q}\surplus(e\mid A)>\frac{k+4}{k+1}\ln k\,w(A)$.

\surplusrecovery*

A positive surplus measures the gain from inserting a set and removing its conflicts after its heaviest neighbor has already been deleted. This suggests grouping pairwise disjoint sets of positive surplus that share the same heaviest neighbor $h$. By inserting these sets together, we incur the cost $w(h)$ only once, while recovering at least the sum of their surpluses. The resulting gain is therefore at least their total surplus minus $w(h)$. We formalize such a group, together with its common heaviest neighbor, as a \emph{star configuration}.

\begin{definition}
     A tuple $(h, S)$, where $h \in A$ and $\emptyset\neq S \subseteq E$, is called a star configuration when the sets in $S$ are disjoint and for all $s \in S$, $\head(s \mid A) = h$. Moreover, we require that for all $s\in S$, $\surplus(s \mid A) > 0$. We call $h$ the \emph{head} of the star and the sets in $S$ its \emph{members}.
\end{definition}

For $S\subseteq E$, write $\mathcal N(S,A)\coloneqq\bigcup_{s\in S}\mathcal N(s,A)$. A star configuration $(h,S)$ defines a swap that inserts $S$ and removes its conflicts from $A$, giving the packing $(A\setminus\mathcal N(S,A))\cup S$. The following claim bounds the change in weight.

\begin{claim}
    \label{clm:star-swap-bound}
    For every star configuration $(h,S)$,
    \begin{align*}
        w((A \setminus \mathcal{N}(S, A)) \cup S) - w(A) \geq \sum_{s \in S} \surplus(s \mid A) - w(h).
    \end{align*}
    Moreover, $(A \setminus \mathcal{N}(S, A)) \cup S$ is a packing.
\end{claim}
\begin{proof}
    First delete $h$ from $A$, at a cost of $w(h)$. Then insert the sets of $S$ one by one, deleting their remaining neighbors. Since the sets in $S$ are disjoint, no insertion deletes an earlier inserted set. Thus, when inserting $s$, the sets deleted form a subset of $\mathcal N(s,A)\setminus\{h\}$, and the gain is at least
    \[
        w(s)-w(\mathcal N(s,A)\setminus\{h\})=\surplus(s\mid A),
    \]
    where the equality uses $\head(s\mid A)=h$ and $\surplus(s\mid A)>0$.
    The resulting family is the packing $(A\setminus\mathcal N(S,A))\cup S$. Summing the insertion gains and subtracting the initial cost $w(h)$ proves the claim.
\end{proof}

By \cref{clm:star-swap-bound}, a star gives an improving swap whenever the total surplus of its members exceeds the weight of its head. To combine several such swaps, we keep all surpluses defined relative to the same reference packing $A$ and use a fractional packing to control their overlap.
For the LP and its rounding, assume $k\in \mathbb{Z}_{\geq 2}$ as in \cref{thm:star-lp-rounding}, and set $\beta\coloneqq\frac{\ln k}{k+1}\in(0,1)$.
Our star configuration packing LP has a variable $x_{h,S}$ for each star configuration $(h,S)$.
Moreover, given a fractional packing, we will later sample a star configuration with probability $\beta$ times its fractional value.
Thus its head costs us at most $\beta x_{h,S}w(h)$ in expectation.
We will show that each member contributes its surplus with probability at least $x_{h,S}/(k+4)$.
This motivates the following objective: after division by $k+4$, it lower-bounds the expected surplus we recover minus the head costs.

\begin{definition}
    Let $\mathcal C$ be the family of star configurations for $A$, and write $v(S)\coloneqq\bigcup_{s\in S}v(s)$ for $S\subseteq E$. The star configuration packing LP is
    \begin{equation}
        \label{LP:star-config-packing}
        \begin{aligned}
            \max\quad &\sum_{(h,S)\in\mathcal C}x_{h,S}
                \left(\sum_{s\in S}\surplus(s\mid A)-(k+4)\beta w(h)\right)\\
            \text{s.t.}\quad
            &\sum_{(h,S)\in\mathcal C:\,u\in v(S)}x_{h,S}\leq1
                &&\forall u\in U,\\
            &x_{h,S}\geq0 &&\forall(h,S)\in\mathcal C.
        \end{aligned}
    \end{equation}
\end{definition}
The constraints limit the total fractional mass using any element to one. In particular, $x_{h,S}\leq1$ for every configuration. Since $0<\beta<1$, the sampling probabilities $\beta x_{h,S}$ are well-defined.

\paragraph{Description of the LP rounding phase}
We round an optimal solution $x$ of \eqref{LP:star-config-packing} using \cref{alg:star-lp-rounding}.

\begin{algorithm}[H]
    \caption{Rounding the star configuration LP}
    \label{alg:star-lp-rounding}
    \DontPrintSemicolon
    \KwIn{A maximal packing $A$ and an optimal solution $x$ to \eqref{LP:star-config-packing}.}
    \KwOut{A packing $B$.}
    Sample each $(h,S)\in\mathcal C$ independently with probability $\beta x_{h,S}$, to get $\mathcal D$\;
    Order $\mathcal D$ uniformly at random as $(h_1,S_1),\ldots,(h_{|\mathcal D|},S_{|\mathcal D|})$\;
    $B\gets A$\;
    \For{$i=1,\ldots,|\mathcal D|$}{
        $B\gets B\setminus\{h_i\}$\;
        \ForEach{$s\in S_i$}{
            \If{$s$ is disjoint from every member of $S_1,\ldots,S_{i-1}$}{
                $B\gets(B\setminus\mathcal N(s,B))\cup\{s\}$\;
            }
        }
    }
    \KwRet{$B$}\;
\end{algorithm}

A member is \emph{accepted} when it passes the disjointness test. This test considers all members of earlier sampled configurations, including not accepted ones. Heads and surpluses are always defined relative to the original packing $A$.

\ifbool{ipcosubm}{%
The proof of \cref{thm:star-lp-rounding} is given in \cref{app:surplus-recovery}.
}{%
\ifbool{ipcosubm}{%
\section{Proof of the surplus-recovery theorem}
\label{app:surplus-recovery}
}{}

\begin{proof}[Proof of \cref{thm:star-lp-rounding}]
    We show that \cref{alg:star-lp-rounding} returns a packing $B$ with the desired expected weight.
    Clearly, the set $B$ returned by the algorithm is a packing.

    To show that it has the desired expected weight, we first account for the gains and losses of the rounding procedure. Each sampled configuration $(h,S)$ incurs a cost of at most $w(h)$ when its head is deleted. When a member $s\in S$ is accepted, it intersects no previously inserted member, and its head has already been deleted. Hence the insertion removes only sets in $\mathcal N(s,A)\setminus\{h\}$ and gains at least
    \[
        w(s)-w(\mathcal N(s,A)\setminus\{h\})=\surplus(s\mid A).
    \]
    This equality uses the positive-surplus requirement for star members.
    Let $Y_{h,S,s}$ indicate that $s$ is accepted when processing $(h,S)$. Summing the changes in weight and taking expectations gives
    \begin{equation}
        \label{eq:B-weight-expected}
        \mathbb E[w(B)]-w(A)
        \geq\sum_{(h,S)\in\mathcal C}
        \left(\sum_{s\in S}\surplus(s\mid A)\mathbb E[Y_{h,S,s}]
             -\beta x_{h,S}w(h)\right).
    \end{equation}

    It remains to show that $\mathbb E[Y_{h,S,s}]\geq x_{h,S}/(k+4)$. Fix a configuration $c\coloneqq(h,S)$ and a member $s\in S$, and write $x_c\coloneqq x_{h,S}$. If $x_c=0$, the bound is immediate, so assume $x_c>0$. Let
    \[
        \mathcal B(c,s)\coloneqq
        \{(g,T)\in\mathcal C\setminus\{c\}\colon v(T)\cap v(s)\neq\emptyset\}
    \]
    be the configurations that can prevent $s$ from being accepted. Since $s$ has at most $k$ elements and the LP allows mass at most one through each element,
    \begin{equation}
        \label{eq:blocker-mass}
        \sum_{d\in\mathcal B(c,s)}x_d
        \leq\sum_{u\in v(s)}\sum_{(g,T) \in \mathcal C:\,u\in v(T)}x_{g,T}
        \leq k.
    \end{equation}
    Thus each member faces at most $k$ units of competing mass.

    To analyze the random order, assign each configuration an independent uniform  time in $[0,1]$, independently of the sampling, and process sampled configurations in increasing time. Condition on $c$ being sampled and having time $t$. Each $d\in\mathcal B(c,s)$ is sampled and precedes $c$ with probability $\beta t x_d$, independently of the others. Therefore,
    \[
        \begin{aligned}
            \mathbb P[Y_{h,S,s}=1\mid c\text{ sampled, time }t]
            &=\prod_{d\in\mathcal B(c,s)}(1-\beta t x_d)\\
            &\geq(1-\beta t)^{\sum_{d\in\mathcal B(c,s)}x_d}
            \geq(1-\beta t)^k.
        \end{aligned}
    \]
    The first inequality follows from concavity of $z\mapsto\log(1-\beta t z)$ on $[0,1]$, which gives $\log(1-\beta t z)\geq z\log(1-\beta t)$; the second uses \eqref{eq:blocker-mass}.
    Integrating over $t$ and multiplying by the sampling probability of $c$ yields
    \begin{align*}
        \mathbb E[Y_{h,S,s}]
        &\geq\beta x_{h,S}\int_0^1(1-\beta t)^k\,dt
        =\frac{1-(1-\beta)^{k+1}}{k+1}\,x_{h,S}\\
        &\geq\frac{k-1}{k(k+1)}\,x_{h,S}
        \geq\frac{x_{h,S}}{k+4}.
    \end{align*}
    Here $(1-\beta)^{k+1}\leq e^{-\beta(k+1)}=1/k$, because $\beta \coloneqq \frac{\ln k}{k+1}$, and the last inequality holds for all $k\geq2$.

    Substituting this bound into \eqref{eq:B-weight-expected} shows that the expected gain is at least $1/(k+4)$ times the LP objective value:
    \begin{equation}
        \label{eq:expected-B-bound}
        \mathbb E[w(B)]-w(A)
        \geq\frac1{k+4}\sum_{(h,S)\in\mathcal C}x_{h,S}
            \left(\sum_{s\in S}\surplus(s\mid A)-(k+4)\beta w(h)\right).
    \end{equation}

    Finally, fix any comparison packing $Q$.
    Discard its zero-surplus sets and group the remaining sets by their heads:
    \[
        Q_h\coloneqq\{e\in Q\colon \surplus(e\mid A)>0,\ \head(e\mid A)=h\}
        \qquad \forall h\in A.
    \]
    Each nonempty group defines a star $(h,Q_h)$. Since $Q$ is a packing, assigning value one to these configurations and zero to all others is feasible for \eqref{LP:star-config-packing}. This solution counts every surplus in $Q$ and pays for each head at most once, so its objective value is at least
    \[
        \sum_{e\in Q}\surplus(e\mid A)-(k+4)\beta w(A).
    \]
    Optimality of $x$ and \eqref{eq:expected-B-bound} consequently give
    \[
        \mathbb E[w(B)]
        \geq w(A)+\frac1{k+4}\sum_{e\in Q}\surplus(e\mid A)
            -\beta w(A).
    \]
    Substituting $\beta=\ln k/(k+1)$ proves the theorem.
\end{proof}

}

\ifbool{ipcosubm}{}{\section{Combining the two phases}
\label{sec:combining-phases}
The guarantees of \cref{thm:phase1,thm:star-lp-rounding} complement each other. The first phase produces a packing whose weight, together with the remaining surplus of the optimum, accounts for the optimum's weight. The second phase recovers a fraction of this surplus while paying for the heads it deletes. We return the heavier of the two packings. The analysis combines the two guarantees so that the surplus term cancels.

\mainapproximation*

\begin{proof}
    If $E=\emptyset$, return $\emptyset$. If $k=1$, select a heaviest set containing each element that occurs in the instance. All sets are singletons, so these choices form an optimal packing. Henceforth, assume $k\geq2$ and $E\neq\emptyset$.
    Run the greedy procedure of \cref{thm:phase1} to obtain a maximal packing $A$, and apply the rounding procedure of \cref{thm:star-lp-rounding} to obtain $B$. Return a heavier packing $C\in\{A,B\}$, breaking ties arbitrarily. Hence, $w(C)=\max\{w(A),w(B)\}$.

    \paragraph{Approximation guarantee}
    We first work with the geometric weights defined in \cref{sec:weight-rounding}. Recall that $\beta=\ln k/(k+1)$, and write 
    \[
        c_k\coloneqq\frac{3(1+r^{-m})}{r+1},
        \qquad
        P\coloneqq\mathbb E\left[\sum_{e\in \OPT}\surplus(e\mid A)\right].
    \]
    Here $\OPT$ is fixed, while the expectation in $P$ is over the first phase. Applying \cref{thm:phase1} with $Q=\OPT$ gives
    \begin{equation}
        \label{eq:combined-phase1}
        c_k w(\OPT)\leq(k+3)\mathbb E[w(A)]+P.
    \end{equation}
    To relate the surplus term to the weight recovered in the second phase, condition on $A$ and apply \cref{thm:star-lp-rounding} with $Q=\OPT$:
    \[
        \mathbb E[w(B)\mid A]
        \geq(1-\beta)w(A)+\frac1{k+4}\sum_{e\in \OPT}\surplus(e\mid A).
    \]
    Taking expectations and rearranging gives
    \begin{equation}
        \label{eq:B-expected-weight}
        P\leq(k+4)\mathbb E[w(B)]-(k+4)(1-\beta)\mathbb E[w(A)].
    \end{equation}
    Substituting into \eqref{eq:combined-phase1} eliminates the surplus term:
    \begin{equation}
        \begin{aligned}
        c_k w(\OPT)
        &\leq(k+4)\mathbb E[w(B)]
            +\bigl((k+4)\beta-1\bigr)\mathbb E[w(A)]
        \ifbool{ipcosubm}{\\ &}{}
        \leq\bigl(k+3+(k+4)\beta\bigr)\mathbb E[w(C)].
        \end{aligned}
        \label{eq:combined-guarantee}
    \end{equation}
    The last inequality uses the pointwise choice of $C$ and the fact that $(k+4)\beta-1\geq0$ for $k\geq2$.

    Thus, keeping the heavier packing pays for both terms in \eqref{eq:combined-guarantee}, regardless of how much surplus remains after the first phase. The resulting approximation factor is
    \[
        \frac{k+3+\frac{k+4}{k+1}\ln k}{c_k}
        =\left(1-\frac1{\sqrt3}\right)k+O(\log k),
    \]
    since $c_k=\frac{3+\sqrt3}{2}(1-O(1/k))$ by \cref{thm:phase1} and $\frac{2}{3+\sqrt3}=1-\frac1{\sqrt3}$.

    \paragraph{Running time}
    Each bucket decreases its upper boundary by a factor of at least $r$. Thus there are at most $\lceil\log_r W\rceil+1=O(1+k\log W)$ buckets, and the first phase makes polynomially many calls to the unweighted subroutine, each taking polynomial time for fixed $k$. 
    The algorithm can also be made strongly polynomial by first discarding sets lighter than a $1/(|E|k)$ fraction of the maximum input weight. As detailed in \cref{sec:derandomization}, rescaling and rounding then gives $W\leq |E|k$, at only an additive $O(1)$ loss in the approximation factor. For fixed $k$, this also bounds the encoding length of the subsequent LP coefficients polynomially in $|E|$, independently of the original weight magnitudes.

    For the second phase, every member of a star intersects its head, and the members are pairwise disjoint. A head has at most $k$ elements, so a star has at most $k$ members. Consequently, there are $O(|E|^{k+1})$ configurations. For fixed $k$, we can enumerate them, solve \eqref{LP:star-config-packing}, and perform the rounding in polynomial time.

    To implement the algorithm with rational arithmetic, we replace the irrational sampling rate $\beta=\ln k/(k+1)$ by a sufficiently close rational number. Standard rounding techniques allow this without changing either phase guarantee. For completeness, \cref{app:rational-implementation} gives the details, including how to sample with polynomially many random bits.
\end{proof}
}

\ifbool{ipcosubm}{}{\section*{Acknowledgements}
We thank Jannis Blauth for many interesting discussions on the topic.
}

\section*{AI acknowledgement}
The main algorithmic ideas were developed without AI assistance. ChatGPT-6 has been used to help complete technical details and improve the exposition of the paper.

\begingroup
\setlength{\emergencystretch}{1em}
\printbibliography
\endgroup

\newpage

\appendix

\ifbool{ipcosubm}{%

}{}

\section{Properties of the bucket distributions}
\label{app:bucket-properties}
We prove the two boundary properties stated in \cref{sec:random-buckets}. The first concerns the position of a fixed grid weight in the full bucketing. The second concerns a continuation from a known boundary.

\begin{proof}[Proof of \cref{lem:idential-dist+expected-ell}]
    The choice of the initial law makes the distance to the next lower boundary stationary as we move down the weight grid. To see this, consider the grid weights $w_j\coloneqq Wr^{-j}$, for integers $j\geq0$, and let $R_j\in\{1,\ldots,m\}$ be the number of grid steps from $w_j$ to the lower boundary of its bucket. Thus that boundary is $w_jr^{-R_j}$. We may continue drawing buckets below the smallest input weight for this argument.

    Initially, $R_0=X_1$ has law $\pi$. If $R_j>1$, moving down one grid step gives $R_{j+1}=R_j-1$. If $R_j=1$, the next grid weight is a bucket boundary and belongs to the bucket below it, so $R_{j+1}$ is a fresh draw from $\nu$. The formulas for $\pi$ and $\nu$ give
    \[
        \pi_i=\pi_{i+1}+\pi_1\nu_i\quad(1\leq i<m),
        \qquad
        \pi_m=\pi_1\nu_m.
    \]
    These identities say precisely that a step preserves the distribution $\pi$. Induction on $j$ therefore shows that every $R_j$ has law $\pi$, proving the distributional claim.

    For a set $e$ with $w(e)=w_j$, it follows that
    \[
        \frac{\mathbb E[\ell(e)]}{w(e)}
        =\sum_{i=1}^m\pi_i r^{-i}
        =\frac{r-1}{1-r^{-m}}\sum_{i=1}^m r^{-2i}
        =\frac{1+r^{-m}}{r+1}.
    \]
    Finally, $r^{m-1}<\sqrt3\leq r^m$ and $r=1+1/k$ imply $r^{-m}=1/\sqrt3+O(1/k)$, giving the stated asymptotic expression.
\end{proof}

\begin{proof}[Proof of \cref{lem:fresh-boundary}]
    Set $a\coloneqq r^{-m}\leq1/\sqrt{3}$. There is a useful equivalent way to generate the next boundary from a current boundary $x$: draw $Z$ uniformly from $[0,1)$, set $V\coloneqq\max\{ax,Zx\}$, and round $V$ down to the grid $\{xr^{-i}\colon i\in\mathbb Z_{\geq0}\}$ to obtain $Y$.
    Indeed, $\mathbb P[Y=xr^{-i}]=r^{-(i-1)}-r^{-i}=\nu_i$ for $1\leq i<m$, and $\mathbb P[Y=xr^{-m}]=r^{-(m-1)}=\nu_m$.

    Fix the target weight $w$. At a boundary $x\geq w$, keep a budget of $x+w$. After drawing $Y$, continue with budget $Y+w$ if $Y\geq w$, and otherwise stop and pay $3Y$. Thus the replacement budget is
    \[
        F_w(Y),\qquad
        F_w(y)\coloneqq
        \begin{cases}
            y+w,&y\geq w,\\
            3y,&y<w.
        \end{cases}
    \]
    We show that its expectation is at most $x+w$. Since $w$ is a grid point, rounding down preserves whether the next boundary is at least $w$, and only decreases the corresponding budget. Hence $F_w(Y)\leq F_w(V)$.
    If $w\leq ax$, then $F_w(V)=V+w\leq x+w$. Otherwise, put $z\coloneqq w/x\in(a,1]$. Accounting for the probability mass at $ax$ and integrating over the remaining uniform choices gives
    \begin{equation*}
        \mathbb E[F_w(V)]
        =3a^2x+\int_a^z3ux\,du+\int_z^1(ux+w)\,du
        =\frac{1+3a^2}{2}\,x+w
        \leq x+w.
    \end{equation*}
    This is where the cutoff $a\leq1/\sqrt{3}$ is used.

    Starting with budget $M+w$, repeat this step until the first boundary below $w$. Each step decreases the boundary by a factor of at least $r$, so there are at most $j+1$ steps. Iterating the conditional expectation bound shows that the expected final payment, $3\mathbb E[\ell_M(w)]$, is at most $M+w$.
\end{proof}

\section{The unweighted packing guarantee}
\label{app:cygan-guarantee}
Any maximal packing for an instance with $k=1$ or $E = \emptyset$ trivially satisfies \cref{thm:cygan-guarantee}; therefore, we assume $E$ is non-empty and $k \geq 2$.
We derive the guarantee from the bounded-pathwidth search of \textcite{cygan_unweighted}. For a packing $A$, an \emph{improving swap} is a packing $T\subseteq E\setminus A$ with $|T|>|\mathcal N(T,A)|$. Its conflict graph is the bipartite graph between $T$ and $\mathcal N(T,A)$, with edges joining intersecting sets. Cygan's algorithm starts from the empty packing and repeatedly applies improving swaps of size $O(\log |E|)$ whose conflict graphs have constant pathwidth. Such swaps can be found in polynomial time for any fixed $k$. The resulting packing $A$ is maximal, because inserting a set with no neighbor in $A$ would be an allowed swap.

Fix any packing $Q\subseteq E$, and let $Q_1$ and $Q_2$ consist of its members with exactly one and two neighbors in $A$, respectively. Two members of $Q_1$ cannot have the same neighbor $e \in A$ as replacing $e$ by both members would be an improving swap of size two and pathwidth one. Thus $|Q_1|\leq|A|$.

The following claim from \cite{cygan_unweighted} also establishes the bound on $Q_2$.

\begin{claim}[Adapted from Claim 3.12 of \cite{cygan_unweighted}]
    \label{clm:adapted_cygan}
    Let $A$ be the output of the bounded-pathwidth local search algorithm of \cite{cygan_unweighted} with parameter $\varepsilon >0$. Fix any packing $Q\subseteq E$, and let $Q_2$ denote the sets in $Q$ with exactly two neighbors in $A$. Then, $|Q_2|\leq (1+\varepsilon)|A|$.\footnote{Claim 3.12 in \cite{cygan_unweighted} is stated for $Q = \OPT$, but its proof uses only disjointness of the sets in $Q$, not optimality.}
\end{claim}

Applying \cref{clm:adapted_cygan} with $\varepsilon = 1$ now proves $|Q_2| \leq 2|A|$, completing the guarantees we need from the unweighted procedure.

\section{Derandomization of the algorithm}\label{sec:derandomization}

We show that the approximation guarantee of \cref{thm:main} can also be achieved deterministically, leading to the statement below.

\begin{theorem}[Main result]
\label{thm:mainDerandomized}
For every fixed $k\in \mathbb{Z}_{\geq 1}$, there is a polynomial-time algorithm for weighted $k$-set packing that returns a packing $C$ satisfying
\[
    w(\OPT)\leq\left(\left(1-\frac1{\sqrt3}\right)k+O(\log k)\right)
         w(C).
\]
\end{theorem}

The empty instance is immediate, and for $k=1$, greedy is optimal.
We therefore assume $E\neq\emptyset$ and $k\geq2$ below.

We eliminate the randomness in the bucketing by enumeration and in the LP rounding by conditional expectations.

\paragraph{Derandomizing the bucketing.}
Set $n\coloneqq|E|$ and $W\coloneqq\max_{e\in E}w(e)$. Discard all sets of weight less than $W/(nk)$. Their total weight is at most $W/k\leq w(\OPT)/k$, so the retained optimum has weight at least $(1-1/k)w(\OPT)$.
Rescale the smallest retained weight to one and round down to powers of $r=1+1/k$ as in \cref{sec:weight-rounding}. The largest rounded weight is at most $nk$.

There are $O(k\log(nk))$ grid levels between $1$ and $nk$. Include also the $m$ levels $r^{-1},\ldots,r^{-m}$, which contain every possible final boundary below $1$. For fixed $k$, there are therefore only $2^{O(1+\log n)}=\mathrm{poly}(n)$ choices of bucket boundaries. Enumerate the resulting bucketings, retaining the largest rounded weight as the initial boundary and a boundary below $1$ as the last. This includes every bucketing in the support of the random construction. Until the final approximation estimate, $w$ denotes the rounded weights of the retained instance.

\paragraph{Rounding the configuration LP.}
Fix the packing $A$ obtained from one bucketing. Use the rational rate $\widehat\beta$ from \cref{app:rational-implementation} in both the LP and its rounding, and compute an optimal LP solution $x$. For $c=(h,S)\in\mathcal C$, write $x_c\coloneqq x_{h,S}$, let $X_c\in\{0,1\}$ indicate that $c$ is sampled, and let $Y_{c,s}\in\{0,1\}$ indicate that $s\in S$ is accepted. Define the real-valued random variable
\[
    Z\coloneqq\sum_{c=(h,S)\in\mathcal C}
        \left(\sum_{s\in S}\surplus(s\mid A)Y_{c,s}-w(h)X_c\right).
\]
The sequential accounting underlying \eqref{eq:B-weight-expected} gives $w(B)\geq w(A)+Z$ for every outcome. Moreover, the proof of \cref{thm:star-lp-rounding} gives, for every packing $Q$,
\[
    \mathbb E[Z]\geq\frac1{k+4}\sum_{e\in Q}\surplus(e\mid A)-\widehat\beta w(A).
\]
It therefore suffices to fix the random choices so that $Z\geq\mathbb E[Z]$.

Equivalently, order all configurations uniformly at random and ignore the unsampled ones. Let $H$ fix a prefix of this order and its sampling indicators, and let $\mathcal R$ be the remaining configurations. For $c=(h,S)\in\mathcal R$ and $s\in S$ unblocked by the members of the sampled prefix, the priority argument in \cref{thm:star-lp-rounding} gives
\[
    \mathbb E[Y_{c,s}\mid H]
    =\widehat\beta x_c\int_0^1
        \prod_{d\in\mathcal R\cap\mathcal B(c,s)}
            (1-\widehat\beta x_d u)\,du.
\]
For a blocked member, this expectation is zero; for a configuration in the prefix, its acceptance indicators are already fixed. Likewise, $\mathbb E[X_c\mid H]$ is either fixed or equals $\widehat\beta x_c$. These formulas determine $\mathbb E[Z\mid H]$. Each integrand is a product of at most $|\mathcal C|$ rational linear factors, so its coefficients and integral have polynomial encoding length and can be computed in polynomial time by multiplication and termwise integration.

At each step, consider all positive-probability extensions fixing the next configuration and its sampling indicator. Choose one maximizing the conditional expectation of $Z$. The current value is an average of these extension values, so it cannot decrease. After $|\mathcal C|$ steps all choices are fixed, and running \cref{alg:star-lp-rounding} with these choices returns a packing $B$ satisfying
\[
    w(B)\geq w(A)+\mathbb E[Z]
    \geq(1-\widehat\beta)w(A)+\frac1{k+4}\sum_{e\in Q}\surplus(e\mid A).
\]
There are at most $2|\mathcal C|$ extensions per step, so this procedure is polynomial for fixed $k$. Since $\widehat\beta\leq\beta$, it gives the guarantee of \cref{thm:star-lp-rounding} deterministically, simultaneously for all $Q$.

\paragraph{Choosing the output.}
For each enumerated bucketing, compute $A$ and the corresponding deterministic $B$, and return a heaviest packing $C$ among all these candidates. Averaging the two phase guarantees over the original random bucketing, as in \cref{sec:combining-phases}, gives
\[
    c_k w(\OPT)
    \leq(k+4)\mathbb E[w(B)]+\bigl((k+4)\beta-1\bigr)\mathbb E[w(A)]
    \leq\bigl(k+3+(k+4)\beta\bigr)w(C),
\]
where $c_k\coloneqq3(1+r^{-m})/(r+1)$ and $\OPT$ is optimal for the rounded retained instance. The last inequality uses $(k+4)\beta-1\geq0$ and the fact that $C$ is at least as heavy as every candidate. This yields the stated factor on the rounded instance. Returning to the original weights multiplies it by at most $r\,k/(k-1)=1+O(1/k)$, adding only $O(1)$ to the approximation factor and proving \cref{thm:mainDerandomized}.

\section{Rational arithmetic and sampling}
\label{app:rational-implementation}
We provide the implementation details deferred from \cref{sec:combining-phases}.
For $k=1$, a greedy algorithm solves the instance optimally and deterministically.
We therefore fix $k\in\mathbb Z_{\geq2}$ below and recall that $r=1+1/k$ and $\beta=\ln k/(k+1)$. The empty instance is immediate, so assume $E\neq\emptyset$.

To implement the second phase with rational arithmetic, use a fixed rational rate $0<\widehat\beta<\beta$ in both the LP objective and the sampling probabilities, chosen so that
\[
    \frac{1-(1-\widehat\beta)^{k+1}}{k+1}\geq\frac1{k+4}.
\]
Such a rate exists because the left-hand side is continuous in the rate and is strictly larger than $1/(k+4)$ at $\beta$, as follows from the probability calculation in the proof of \cref{thm:star-lp-rounding}. For fixed $k$, the rate can be computed once, independently of the instance. The same proof then gives \cref{thm:star-lp-rounding} with a head cost $\widehat\beta w(A)\leq\beta w(A)$, so the approximation analysis in \cref{sec:combining-phases} still applies. All LP data have polynomial encoding length, as do the rational bucket probabilities in \eqref{eq:nu-def} and \eqref{eq:pi-def}.

To obtain a polynomial worst-case running time using fair random bits, we can truncate the rejection samplers without changing either phase guarantee. In either phase, rescale the minimum weight to one and set $n\coloneqq|E|$ and $W\coloneqq\max_{e\in E}w(e)$, so all working weights lie in $[1,W]$. The free-set analysis in the first phase actually gives the stronger coefficient $k+3/r$ in place of $k+3$. Since every maximal packing of a nonempty instance has weight at least one, this leaves slack at least $3(1-1/r)=3/(k+1)$ in the stated bound. Truncate the first-phase samplers with total failure probability at most
\[
    \varepsilon_1\coloneqq\frac{3}{(k+1)(k+4)nW},
\]
returning any maximal packing on failure. For every fixed comparison packing $Q$, the random quantity $(k+3)w(A)+\sum_{e\in Q}\surplus(e\mid A)$ lies in $[0,(k+4)nW]$. Coupling the truncated and exact executions therefore changes its expectation by at most $\varepsilon_1(k+4)nW$, which is absorbed by the slack. Thus \cref{thm:phase1} is preserved.

For the second phase, fix a rational $d\in(0,\beta-\widehat\beta)$ and truncate the samplers with total failure probability at most $\varepsilon_2\coloneqq d/(nW)$, returning $B=A$ on failure. The expected weight loss is at most $\varepsilon_2nW=d$, which is absorbed by the saved head cost $(\beta-\widehat\beta)w(A)$. Finally, each rational draw can be generated by rejection sampling with success probability at least $1/2$ per trial. If $N$ bounds the number of draws in a phase, allowing $\lceil\log_2(N/\varepsilon_i)\rceil$ trials per draw gives total failure probability at most $\varepsilon_i$. Both $N$ and the required number of bits are polynomial in the input length for fixed $k$, including for the random ordering of sampled configurations. This proves the claimed running-time bounds.

\end{document}